\documentclass[12pt]{article}
\pdfoutput=1
\usepackage{color,graphicx,amsfonts,amsmath,amssymb,mathrsfs, comment}
\usepackage{citesort}
\usepackage{url}

\newcommand{\bq}{\begin{quote}}
\newcommand{\be}{\begin{equation}}
\newcommand{\ee}{\end{equation}}
\newcommand{\psisq}{|\psi|^2}
\newcommand{\eq}{\end{quote}}

\newcommand{\bc}{\begin{center}}
\newcommand{\ec}{\end{center}}
\definecolor{gray}{gray}{.4} 
\definecolor{lightgray}{gray}{.7}

\begin{document}

\title{Bohmian Mechanics and Quantum Information\footnote{Dedicated to
    Jeffrey Bub on the occasion of his 65th birthday.}}

\author{Sheldon Goldstein \\
Departments of Mathematics, Physics,  and Philosophy -- Hill Center \\
Rutgers, The State University of New Jersey \\
110 Frelinghuysen Road \\
Piscataway, NJ 08854-8019, USA \\
E-mail: oldstein@math.rutgers.edu}

\date{July 25, 2007}

\maketitle

\begin{abstract}

Many recent results suggest that quantum theory is about
information, and that quantum theory is best understood as arising from
principles concerning information and information processing. At the same time,
by far the simplest version of quantum mechanics, Bohmian mechanics, is
concerned, not with information but with the behavior of an objective
microscopic reality given by particles and their positions.  What I would
like to do here is to examine whether, and to what extent, the importance of
information, observation, and the like in quantum theory can be understood
from a Bohmian perspective. I would like to explore the hypothesis that the
idea that information plays a special role in physics naturally emerges in
a Bohmian universe.

\end{abstract}

\bibliographystyle{unsrt}

\section{Introduction: The Status of the  Wave Function}
 Few people have struggled as long and as hard with the foundations of
 quantum mechanics as Jeffrey Bub, and even fewer have done so with as much
 seriousness, honesty, and gentleness.  Jeffrey has in fact explored more
 or less all approaches to the interpretation of quantum mechanics, and has
 made seminal contributions to most of them. I am indeed pleased---and
 honored---to have been invited to contribute to this volume in honor of
 Jeffrey, and thank the organizers for having done so. 

 The question animating the foundations quantum mechanics, for Jeffrey and
 everyone else in the field, is this: What is the nature of the reality, if
 any, that lies behind the quantum mathematics? Now the reality issue, in
 quantum mechanics and in general, is difficult and controversial. Here are
 two quotations that, while expressing in very different ways the subtleties
 involved, nonetheless get to the core of the problem---and its
 solution---and indeed express pretty much the same thing:
\bq What if everything is an illusion and nothing exists? In that case, I
definitely overpaid for my carpet. ({\em Woody Allen})

I did not grow up in the Kantian tradition, but came to understand the
truly valuable which is to be found in his doctrine, alongside of errors
which today are quite obvious, only quite late. It is contained in the
sentence: ``The real is not given to us, but put to us (by way of a
riddle).'' This obviously means: There is such a thing as a conceptual
construction for the grasping of the inter-personal, the authority of which
lies purely in its validation. This conceptual construction refers
precisely to the ``real'' (by definition), and every further question
concerning ``the nature of the real'' appears empty. ({\em Albert
  Einstein})
\eq 
Many readers will perhaps find what Einstein \cite[page 680]{einstein49} says here
too realistic. To others it will no doubt seem too positivistic. For me,
however, it is right on target.

Perhaps the most puzzling object in quantum mechanics is the wave function,
concerning which many basic questions can be asked:

\begin{itemize}
\item Is it subjective or objective?
\item Does it merely represent information or does it describe an observer independent reality?
\item If it  is objective, does it represent a  concrete material sort of
  reality, or does it somehow have an entirely different and perhaps novel
  nature?
\item What's the deal with collapse?
\end{itemize}
There seems to be little agreement about the answers to these questions.
But we can at least all agree that one of the following crudely expressed
possibilities for the wave function must be correct:

\begin{enumerate}
\item The wave function is everything.
\item The wave function is something (but not everything).
\item The wave function is nothing.
\end{enumerate}
The second possibility, which amounts to the suggestion that there are, in
addition to the wave function, what are often called hidden variables, is
regarded by the physics community as the least acceptable and most
implausible of these three possibilities---the very terminology ``hidden
variables'' points to the unease.  This is interesting since it would also
seem to be the most modest of the three. 

The third possibility is best associated with the view that the wave
function of a system is merely a representation of our information about
that system. However, supporters of this view very often also seem to
subscribe to the first possibility as well, at least insofar as microscopic
reality is concerned. (I shall argue later that also (2) and (3) are not as
incompatible as they seem to be.) But here we should recall the words of
Bell \cite[page 201]{bell87}, concerning the theories that reject (1) in favor of (2):
\bq
  Absurdly, such theories are known as ``hidden variable'' theories.
  Absurdly, for there it is not in the wavefunction that one finds an image
  of the visible world, and the results of experiments, but in the
  complementary ``hidden''(!) variables. {Of course the extra
    variables are not confined to the visible ``macroscopic'' scale.  For
    no sharp definition of such a scale could be made. The ``microscopic''
    aspect of the complementary variables is indeed hidden from us. But to
    admit things not visible to the gross creatures that we are is, in my
    opinion, to show a decent humility, and not just a lamentable addiction
    to metaphysics.} In any case, the most hidden of all variables, in the
    pilot wave picture, is the wavefunction, which manifests itself to us
    only by its influence on the complementary variables. 
\eq

The idea that the wave function merely represents information, and does not
describe an objective state of affairs, raises many questions and problems:
\begin{itemize}
\item Information about what?
\item What about quantum interference? How can the terms of a quantum
  superposition interfere with each other, producing an observable
  interference pattern, if such a superposition is just an expression of
  our ignorance?
\item The problem of vagueness: Quantum mechanics is supposed to be a
  fundamental physical theory. As such it should be precise. But if it is
  fundamentally about information, then it is presumably concerned directly
  either with mental events or, more likely, with the behavior of macroscopic
  variables. But the notion of the macroscopic is intrinsically vague.
\item Simple physical laws are to be expected, if at all,
  at the most fundamental level---of the basic microscopic entities---and
  that messy complications should arise at the level of larger complex
  systems. It is only at this level that talk of information, as
  opposed to microscopic reality,  can become appropriate.
\item The very form of the Hamiltonian and wave function strongly points to a
  microscopic level of description.
\item There is a widespread belief that large things are built out of
  small ones, and that to understand even the large we need to understand
  the small.  
\end{itemize}
Nonetheless, many arguments suggest that quantum mechanics is about
information, or that the wave function represents information. (This
suggestion is usually accompanied by the claim that if you ask for
more---if you try to regard quantum mechanics or the wave function as
describing an objective microscopic reality---you get into trouble.) I
don't want to directly criticize these here. Rather I want to observe that
Bohmian mechanics, the simplest version of quantum mechanics---discovered
by Louis de Broglie \cite[page 119]{debroglie28} in 1927 and rediscovered by
Jeffrey's mentor David Bohm
\cite{bohm521}
in 1952---does do more, and thus I want to try to understand
how, from the perspective of Bohmian mechanics, the informational aspect of
the wave function or the quantum state can seem natural. I wish to discuss
in particular the following three informational aspects of the wave
function in Bohmian mechanics: \begin{itemize}
\item The wave function  as a property of the environment.
\item The wave function  as providing the best possible information about
  the system (given by $|\psi|^2$).
\item The wave function as nomological.
\end{itemize}
I note as well that Bohm and Hiley \cite{bohm93} wrote of the wave function
as ``active information.''

Before proceeding to the description of Bohmian mechanics, I would like to
recall the conventional wisdom on the subject. So here are three typical
recent statements about hidden variables and the like, the second from a
very popular textbook on quantum mechanics. The reader should bear these in
mind when reading about Bohmian mechanics. In particular, he or she should
contrast the simplicity of Bohmian mechanics with the complexity,
implausibility or artificiality suggested by the quotations.

\bq
Thus, unless one allows the existence of contextual hidden variables with
very strange mutual influences, one has to abandon them---and, by
extension, `realism' in quantum physics---altogether. ({\em Gregor Weihs}
  \cite[The truth about reality]{weihs})
\eq

\bq
  Over the years, a number of hidden variable theories have been proposed,
  to supplement q.m.; they tend to be {cumbersome and
    implausible}, but never mind--until 1964 the program seemed eminently
  worth pursuing.  But in that year J.S. Bell proved that
\emph{any} local  {hidden variable} is
\emph{incompatible} with   
quantum mechanics.\footnote{I shall not address in this paper the issue of
  nonlocality. But what is misleading about the last sentence is its
  suggestion that the source of the incompatibility is the assumption of
  hidden variables \cite[pages 143 and 150]{bell87}. What Bell in fact
  showed is that the source of the difficulty is the assumption of
  locality. He showed that quantum theory is intrinsically nonlocal, and
  that this nonlocality can't be eliminated by the incorporation of hidden
  variables.} ({\em D.J.\ Griffiths} \cite[page 423]{griffiths})
\eq
\bq
Attempts have been made by Broglie, David Bohm, and others to construct
theories based on hidden variables, but the theories are {very complicated
  and contrived}. For example, the electron would definitely have to go
through only one slit in the two-slit experiment. To explain that
interference occurs only when the other slit is open, it is necessary to
postulate a special force on the electron which exists only when that slit
is open. Such artificial additions make hidden variable theories
unattractive, and there is little support for them among physicists. ({\em
  Encyclopedia Britannica} \cite{britannica})
\eq

\section{Bohmian Mechanics}

In Bohmian mechanics the state of an $N$-particle system is given by its
wave function $\psi=\psi(\mathbf q_{_1}, \dots, \mathbf q_{_N})=\psi(q)$
together with the positions $ \mathbf Q_{_1}, \dots, \mathbf Q_{_N}$, forming
the configuration $Q$, of its particles. The latter define the {\em
  primitive ontology} (PO) \cite{allori07} of Bohmian mechanics,
what the theory is fundamentally about. The wave function, in contrast, is
not part of the PO of the theory, though that should not be taken to
suggest that it is not objective or real. It plays a crucial role in
expressing the dynamics for the particles, via a first-order differential
equation of motion for the configuration $Q$, of the form $dQ/dt=v^{\psi}(Q)$.

The defining equations of Bohmian mechanics are Schr\"odinger's equation
\be\label{SE}
 i\hbar\frac{\partial \psi}{\partial t} = H\psi \,,
\ee
where  
\begin{equation}\label{H} 
H=-\sum_{k=1}^N\frac{\hbar^2}{2m_k}\nabla^2_k+V,
\end{equation} 
for the wave function, and the {\em guiding equation} 
\begin{equation}\label{g}
  \frac{d\mathbf Q_k}{dt}=\frac{\hbar}{m_k}
  \mbox{Im} \frac{\psi^{*}\mathbf{\nabla}_k \psi}{\psi^{*}\psi}(\mathbf
  Q_{_1}\ldots,\mathbf Q_{_N}) 
\end{equation} 
for the configuration. In the Hamiltonian \eqref{H} the $m_k$ are of course
the masses of the particles and $V=V(q)$ is the potential energy function. For
particles with spin, the products involving $\psi$ in the numerator and the
denominator of \eqref{g} should be understood as spinor inner products, and
when magnetic fields are presents, the $ {\nabla}_k$ in
\eqref{H} and  \eqref{g} should be understood as a covariant derivative,
involving the vector potential $\boldsymbol A=\boldsymbol A(\mathbf q_{_k})$.

For particles without spin, the $\psi^*$ in the guiding equation \eqref{g}
cancels, and the equation assumes the more familiar form
\begin{equation}\label{S}
\frac{d\mathbf Q_k}{dt}=\frac{\mathbf{\nabla}_k S}{m_k}
\end{equation} 
where $S$ arises from the polar decomposition $\psi=Re^{iS/\hbar}$ with
 $S$ real and $R\geq 0$. Equation \eqref{g} however has two advantages: 
\begin{itemize}
\item It  is explicitly of the form $\frac{J_k}{\rho}$ with $J_k$ the quantum
  probability current and $\rho=\psi^*\psi=\psisq$ the quantum probability
  density, a fact of great importance for the statistical implications of
  Bohmian mechanics.
\item With the guiding equation in this form, Bohmian
  mechanics applies without further ado also to particles with spin; in
  particular there is no need to associate any additional discrete spin
  degrees of freedom with the particles---the fact that the wave function is
  spinor valued entirely takes care of the phenomenon of spin.
\end {itemize}

A surprising and striking fact about Bohmian mechanics is its simplicity
and obviousness. Indeed, given Schr\"odinger's equation, from which one
immediately extracts $J$ and $\rho$, related classically by $J=\rho v$, it
takes little imagination when looking for an equation of motion for the
positions of the particles in quantum mechanics to consider the possibility
that $v=J/\rho$, which is precisely \eqref{g}. 

But even without having arrived at Schr\"odinger's equation, or parallel
with doing so, we could easily guess the guiding equation \eqref{S} for
particles without spin: The de Broglie relation ${\mathbf p} = \hbar {\mathbf
  k}$ is a remarkable and mysterious distillation of the experimental facts
associated with the beginnings of quantum theory. This relation, connecting
a particle property, the momentum ${\mathbf p}=m{\mathbf v}$, with a wave
property, the wave vector $ {\mathbf k}$, immediately yields Schr\"odinger's
equation, giving the time evolution for $\psi$, as the simplest wave
equation that reflects this relationship. This is completely standard and
very simple.  Even simpler, but not at all standard, is the connection
between the de Broglie relation and the guiding equation, giving the time
evolution for $Q$: The de Broglie relation says that the velocity of a
particle should be the ratio of $\hbar {\mathbf k}$ to the mass of the
particle. But the wave vector ${\mathbf k}$ is defined for only for a plane
wave. For a general wave $\psi$, the obvious generalization of $ {\mathbf k}$
is the local wave vector $\mathbf{\nabla}S({\mathbf q})/\hbar$, and with this
choice the de Broglie relation becomes the guiding equation $dQ/dt = \nabla
S / m$.

\section{The Implications of Bohmian Mechanics}
That a theory is simple and obvious doesn't make it right. And in
the case of Bohmian mechanics this fact suggests in the strongest possible
terms that it must be wrong. If something so simple could account
for quantum phenomena, it seems extremely unlikely that it would have been
ignored or dismissed by almost the entire physics community for so many
decades---and in favor of alternatives which seem at best far more radical.

Of course, one can see at a glance, see Fig. 1,
\begin{figure}[h]\label{slit}
\bc\includegraphics[width=13cm,height=9cm]{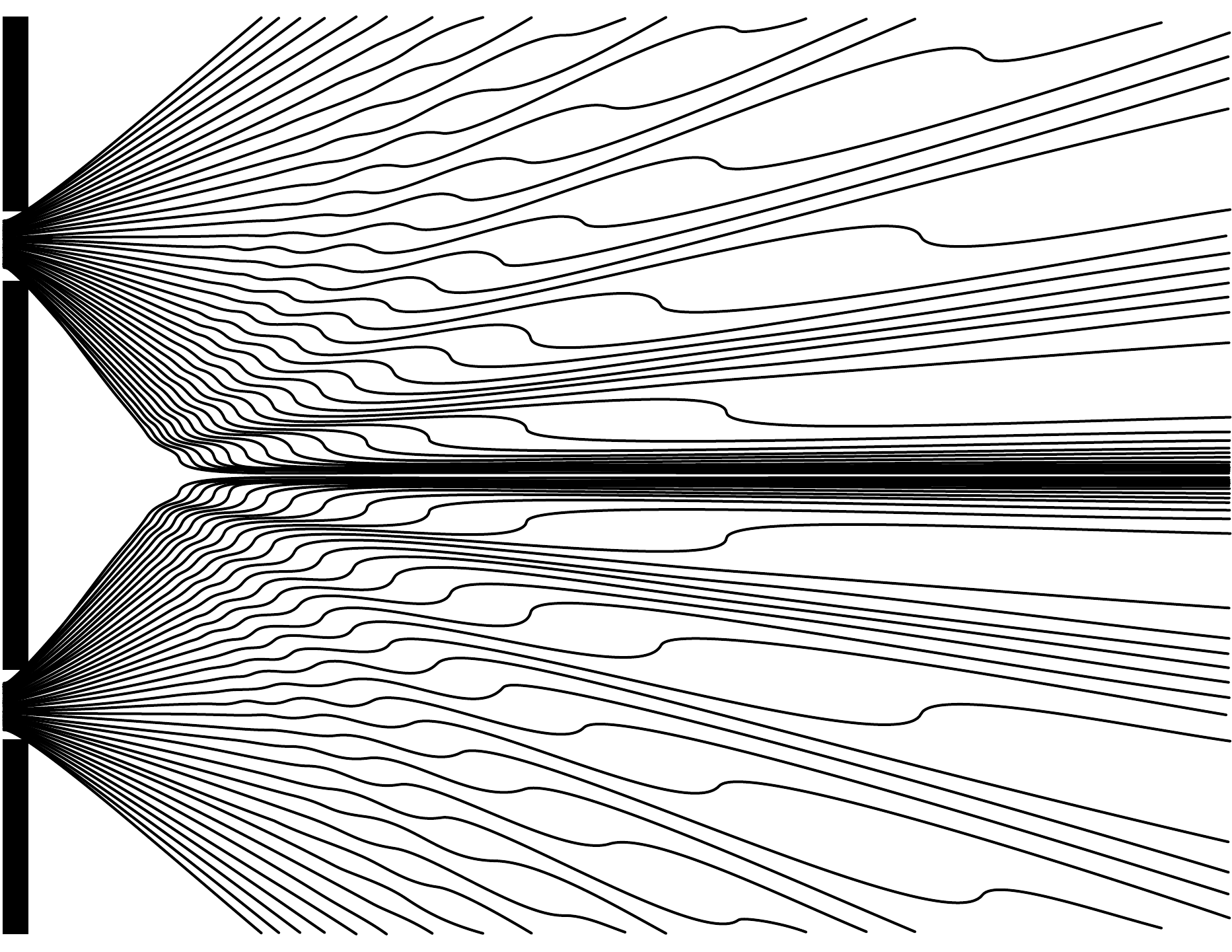}\ec
\caption{An ensemble of trajectories for the two-slit experiment, uniform
in the slits. (Drawn by G. Bauer from \cite{philippidis79}.)}
\end{figure}
that Bohmian mechanics seems to handle one of the characteristic mysteries
of quantum mechanics, the two-slit experiment, quite well. One sees in Fig.
1, in this ensemble of Bohmian trajectories with an approximately uniform
distribution of initial positions in the slits, how an interference-like
profile in the pattern of trajectories develops after the parts of the wave
function emerging from the upper and lower slits begin to overlap.

This of course does not prove that Bohmian mechanics makes the same
quantitative predictions for the two-slit experiment---let alone the same
predictions for all quantum experiments---as orthodox quantum theory, but
it in fact does. Bohmian mechanics is entirely empirically equivalent to
orthodox quantum theory, as least insofar as the latter is unambiguous.
This was basically shown by Bohm in his first papers \cite{bohm521,bohm522}
on the subject, modulo the status in Bohmian mechanics of the Born
probability formula $\rho=\psisq$. That issue was addressed in \cite{durr92}
and is now completely understood. In particular, as a consequence of
Bohmian mechanics one obtains the following:
\begin{enumerate}
\item familiar (macroscopic) reality
\item formal scattering theory \cite{durr061}
\item operators as observables \cite{bohm521,bohm522,durr041}
\item quantum randomness \cite{durr92} 
\item absolute uncertainty \cite{durr92}
\item the wave function  of a (sub)system \cite{durr92}
\item collapse of the wave packet \cite{durr041}
\end{enumerate}

Concerning these, a few comments. Since macroscopic objects are normally
regarded as built out of microscopic constituents, which of course could be
point particles, there can be no problem of macroscopic reality per se in
Bohmian mechanics. Less obvious, but reasonably clear \cite{allori02}, is the
fact that in a Bohmian universe macroscopic objects behave classically, for
example moving according to Newton's equations of motion as appropriate.

The picture of what occurs in a Bohmian scattering experiment, in which
particles are directed at a target---or at each other---with which they
collide and scatter in an apparently random direction, is exactly the
picture that an experimentalist has in mind. Moreover, the additional
structure (actual particles!)  afforded by Bohmian mechanics allows one to
considerably sharpen traditional scattering theory both conceptually and
indeed mathematically.

It should be noted that operators as observables play no role whatsoever in
the formulation of Bohmian mechanics. In fact the only quantum operator
that appears in the defining equations of Bohmian mechanics is the
Hamiltonian $H$, but merely as part of an evolution equation. Nonetheless,
it turns out that operators on Hilbert space are exactly the right
mathematical objects to provide a compact representation of the statistics
for the results of experiments in a Bohmian universe.

I wish to focus here in more detail on items 4--7, which are quite relevant
to my main concern here, the informational aspects of the wave function in
Bohmian mechanics, and which, as it turns out, come together as a package.
For example, the statistical properties of the collapse of the wave packet
depend upon quantum randomness. It should be noted that the claim that the
collapse of the wave packet is an implication of Bohmian mechanics should
seem paradoxical, since Schr\"odinger's equation is an absolute equation of
Bohmian mechanics, never to be violated---unlike the situation in orthodox
quantum theory.

A crucial ingredient in the emergence of quantum randomness is the {\em
  equivariance} of the probability distribution on configuration space
given by $\rho^\psi=\psisq$. This means that 
\be \left(\rho^\psi\right)_t=\rho^{\psi_t}\ee
where on the left we have the evolution of the probability distribution
under the Bohmian flow \eqref{g} and on the right the probability
distribution associated with the evolved wave function $\psi_t$. That this
is so for \be\rho^\psi(q) = |\psi(q)|^2\ee is, by \eqref{g}, equivalent to
the quantum continuity equation.  The equivariance of $\rho^\psi=\psisq$
means that {\em if $\rho_{t_0}(q) = |\psi_{t_0}(q)|^2 $ at some time $t_0$
  then $\rho_t(q) = |\psi_t(q)|^2$ for all $t$.} It says that
Schr\"odinger's equation and the guiding equation are compatible modulo
$\rho=\psisq$.

The upshot of a long analysis \cite{durr92} that begins with the
equivariance of $\rho^\psi=\psisq$ is that the {\it quantum equilibrium\/}
given by $\rho_{qe}(q) = |\psi(q)|^2$ has a status very much the same as
that of {\it thermodynamic equilibrium\/}, described in part by the
Maxwellian velocity distribution $\rho_{eq}(\mathbf v)\propto e^{-
  \frac{1}{2}m{\mathbf v}^2/kT}$ for the molecules of a gas in a box in
equilibrium at temperature $T$. It has recently been shown \cite{goldstein07} that
quantum equilibrium is unique. More precisely, it has been shown that
$|\psi(q)|^2$ is the only equivariant distribution that is, in a
natural sense, a local functional of the wave function.

In order to grasp the meaning of quantum equilibrium, to appreciate the
physical significance $\rho_{qe}(q) = |\psi(q)|^2$, one must first address
this question: in a Bohmian universe with wave function $\Psi$, what is to
be meant by the wave function $\psi$ of a subsystem of that universe? 

\section{The Wave Function of a Subsystem}

Consider a Bohmian universe. This is completely described by its wave
function $\Psi$, the wave function of the universe, and its configuration
$Q$. Given an initial condition $\Psi_0$ and $Q_0$ for this universe, the
equations of motion \eqref{SE} and \eqref{g} determine the trajectories of
all particles throughout all of time and hence everything that could be
regarded as physical in this universe. However, we are rarely concerned
with the entire universe. What we normally deal with in physics is the
behavior of a system that is a subsystem of the universe, usually a small
one such as a specific hydrogen atom.

It is important to appreciate that a subsystem of a Bohmian universe is not
ipso facto itself a Bohmian system. After all, the behavior of a
part is entirely determined by the behavior of the whole, so we are not
free to stipulate the behavior of a subsystem of a Bohmian universe, in
particular that it be Bohmian, having its own wave function that
determines the motion of its configuration in a Bohmian way. Nonetheless,
there is a rather obvious candidate for the wave function of a subsystem,
at least for a universe of spinless particles, and this obvious candidate
behaves in exactly the manner that one should expect for a quantum
mechanical wave function. (For particles with spin the situation is a
little more complicated, so I will confine the presentation here to the
case of spinless particles.) This is the conditional wave function, to
which I now turn. 

Fig. 2 depicts a system corresponding to particles in a
certain region (at a given time), a region surrounded by the rest of the
universe, in which are contained (at that time) the particles of what we'll
call the {\em environment} of the system.
\begin{figure}[h]\label{system}
\bc\includegraphics[width=12cm]{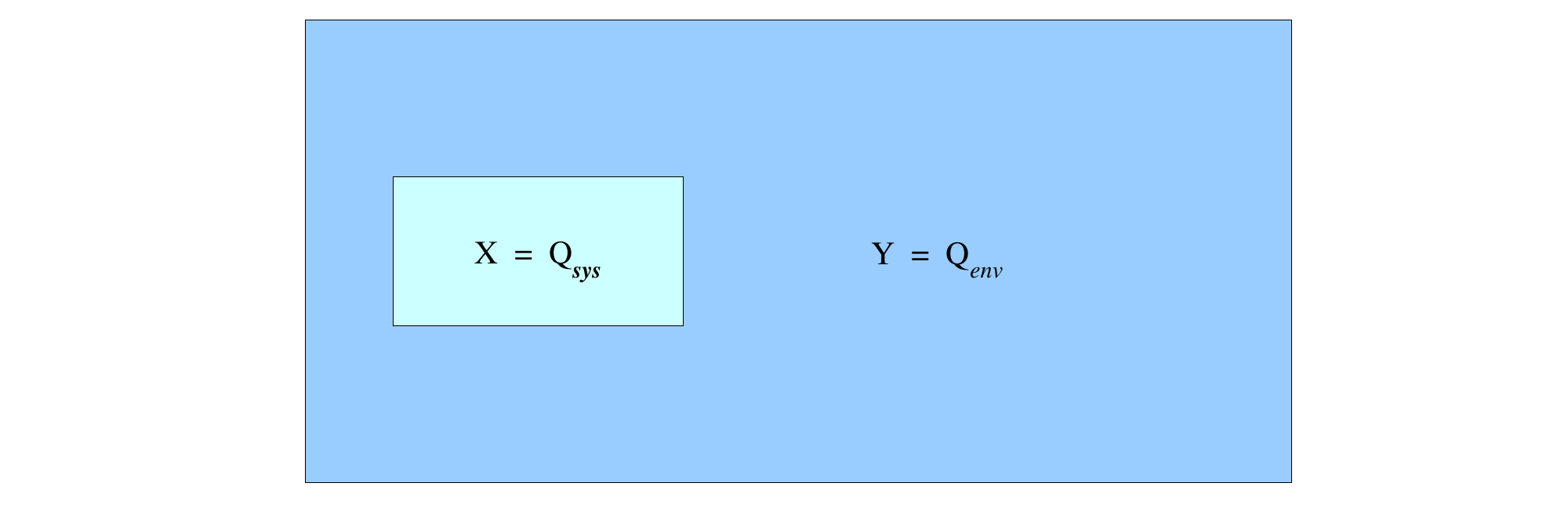}\ec
\caption{A subsystem of a Bohmian universe}
\end{figure}
Corresponding to this system we we have a splitting
$Q=(Q_{sys},Q_{env})=(X,Y)$ of the configuration of the universe into the
configurations of system, $Q_{sys}=X$, and environment, $Q_{env}=Y$.

The wave function $\psi$ of the system must be
constructed from $\Psi$, $X$, and $Y$, since these provide the complete
description of our Bohmian universe (at a given time). The right
construction is the following: The wave function $\psi$ of the system, its
{\em conditional wave function } is given by
\be\label{cwf}
 \psi(x)=\Psi(x,Y).
\ee
Putting in the explicit time dependence, we have that 
\be\label{cwft}
\psi_t(x)=\Psi_t(x,Y_t).
\ee
Here $Y_t$ is the evolving configuration of the environment, corresponding
to the configuration $Q_t=(X_t,Y_t)$, which evolves according to the guiding
equation \eqref{g} (for the universe, with $\Psi$ instead of $\psi$).

Note that the conditional wave function, as given in \eqref{cwf} and
\eqref{cwft}, need not be normalized. In fact these equations should be
understood projectively, as defining a ray in the Hilbert space for the
system, with wave functions related by a (nonzero) constant factor regarded as
equivalent. Of course it is important in probability formulas involving the
wave function that it be normalized. In any such formulas it will be
assumed that this has been done. 

Because of the double time dependence in \eqref{cwft}, the conditional wave
function $\psi_t$ evolves in a complicated way, and need not obey
Schr\"odinger's equation for the system. Nonetheless, it can be shown
\cite{durr92} that
it does evolve according to Schr\"odinger's equation when the system is
suitably decoupled from its environment. While most readers are probably
prepared to accept this, since they are quite accustomed to wave functions
obeying Schr\"odinger's equation, that this is so is a bit delicate. What
is really easy to see, but what most readers are likely to resist, is the
fact, derived in the next subsection, that this wave function collapses
according to the usual textbook rules when the system interacts with its
environment in the usual measurement-like way. 

But before turning to that we should pause to examine the construction
\eqref{cwf} of the conditional wave function  a little more closely. We would expect a property of a system
to correspond to a function of its basic variables---e.g., of its
configuration.  Note, however, that $\psi$ is a function of the
configuration $Y$ of the environment---like a property of the environment!
And to the extent that we come to know $\psi$, that property of the
environment can be identified with what we would tend to regard as
information about the system---so that it is perhaps only a bit of a
stretch to say that $\psi$ represents, or is, our information about the
system. (But it is still a stretch.)

\subsection{Collapse of the Wave Packet}

Consider a quantum observable for the system, given by a self-adjoint
operator  $A$ on its Hilbert space. For simplicity we assume that $A$ has
non-degenerate point spectrum, with normalized eigenstates
$\psi_{\alpha}(x)=|A=\alpha\rangle$, $\|\psi_{\alpha}\|=1$,
\be
A\psi_{\alpha}(x)= \alpha \psi_{\alpha}(x)
\ee
corresponding to the eigenvalues $\alpha$. According to standard quantum
measurement theory, what is called an ideal measurement of $A$ is
implemented by having the system interact with its environment in a
suitable way. (To avoid complications we shall assume here that this
environment consists of a suitable apparatus, and that the rest of the
environment of the system can be ignored---for the the wave function
evolution, for the evolution of the configuration of system and apparatus,
and for the definition of the conditional wave function of the system. Thus
in what follows the configuration of the apparatus will be identified with
the configuration $Y$ of the environment of the system.)

The measurement begins, say, at time 0, with the initial (``ready'') state of
the apparatus given by a wave function $\Phi_0(y)$, and ends at time
$t$.  The interaction is such that when the state of the system is
initially $\psi_{\alpha}$ it produces a normalized apparatus state
$\Phi_{\alpha}(y)=|A_{app}=\alpha\rangle$, $\|\Phi_{\alpha}\|=1$, that
registers that the value found for $A$ is $\alpha$ without having affected
the state of the system,
\be
 \psi_{\alpha}(x)\Phi_0(y)\stackrel{t}{\rightarrow}\psi_{\alpha}(x)\Phi_{\alpha}(y).
\ee
Here $\stackrel{t}{\rightarrow}$ indicates the unitary evolution induced by
the interaction. If the measurement is to provide useful information, the
apparatus states must be noticeably different, corresponding, say,
to a pointer on the apparatus pointing in different directions. We thus
have that the $\Phi_{\alpha}$ have disjoint supports in the configuration
space for the environment,
\be
\text{supp}(\Phi_{\alpha})\cap\text{supp} (\Phi_{\beta})=\emptyset,\  \alpha\neq\beta.
\ee

Now suppose that the system is initially, not in an eigenstate of $A$, but in
a general state, given by a superposition
\be
\psi(x)=\sum_{\alpha}c_{\alpha}\psi_{\alpha}(x).
\ee
We then have, by the linearity of the unitary evolution, that
\be
\Psi_0(x,y)=\psi(x)\Phi_0(y)\stackrel{t}{\rightarrow}\Psi_t(x,y)=\sum_{\alpha}c_{\alpha}\psi_{\alpha}(x)\Phi_{\alpha}(y),
\ee
so that the final wave function $\Psi_t$ of system and apparatus is itself
a superposition. The fact that the pointer ends up pointing in a definite
direction, even a random one, is not discernible in this final wave
function. Insofar as orthodox quantum theory is concerned, we've arrived at
the measurement problem.

However, insofar as Bohmian mechanics is concerned, we have no such
problem, because in Bohmian mechanics particles always have positions and
pointers, which are made of particles, always point---in a direction
determined by the final configuration $Y_t$ of the apparatus. Moreover, in
Bohmian mechanics we find that the state of the system is transformed in
exactly the manner prescribed by textbook quantum theory.

We have---and this is no surprise---that the initial wave function of the
system is  
\be
\psi_0(x)=\Psi_0(x,Y_0)=\psi(x)\Phi_0(Y_0) \stackrel{p}{=}\psi(x).
\ee
And for the final wave function of the system we have that 
\be
\psi_t(x)=\Psi_t(x,Y_t)=\sum_{\alpha}c_{\alpha}\psi_{\alpha}(x)\Phi_{\alpha}(Y_t)=c_a\psi_a(x)\Phi_a(Y_t)\stackrel{p}{=}\psi_a(x)
\ee
when $Y_t\in\text{supp}(\Phi_a)$. Here the $\stackrel{p}{=}$ refers to
projective  equality, and reminds us that the wave function is to be regarded
projectively in Bohmian mechanics. 

Thus in Bohmian mechanics the effect of ideal quantum measurement on the
wave function of a system is to produce the transition
\be
\psi(x)\rightarrow\psi_a(x) \quad \text{with probability }p_a,
\ee   
where $p_a$ is the probability that $ Y_t\in\text{supp}(\Phi_a)$, i.e., that
the value $a$ is registered. Assuming the {\em quantum equilibrium
  hypothesis}, that when a system has wave function $\Psi$ its
configuration is random, with distribution $|\Psi(q)|^2$, we find, by
integrating $|\Psi_t(x,y)|^2$ over $\text{supp}(\Phi_a)$, that $p_a=|c_a|^2
$, the usual textbook formula for the probability of the result of the
measurement.

\subsection{The Fundamental Conditional Probability Formula}

The analysis just given suggests---and it is indeed the case
\cite{durr92,durr041}---that Bohmian mechanics is empirically equivalent to orthodox
quantum theory provided we accept the quantum equilibrium hypothesis.  But
that the quantum equilibrium hypothesis is true, and even what exactly it
means, is a tricky matter, requiring a careful analysis \cite{durr92}
involving typicality that I shall not delve into here. Rather, I shall
focus on a simple but important ingredient of that analysis, a probability
formula strongly suggesting a connection, if not quite an identification,
between the wave function of a system and our information about that
system. 

This {\em fundamental conditional probability formula} is the
following:
\be\label{fcpt}
P(X_t\in dx\,|\,Y_t)=|\psi_t(x)|^2dx. 
\ee
Here $P$ is the probability distribution on universal Bohmian trajectories
arising from the distribution $|\Psi_0|^2$ on the initial configuration of
the universe, with the initial time $t=0$ the time of the big bang, or
shortly thereafter. Of course, by the equivariance of the $|\Psi|^2$
distribution, $|\Psi_t|^2$ at any other time $t$ would define the same
distribution on trajectories. The formula says that the conditional
distribution of the configuration $X_t$ of the system at time $t$, given
the configuration $Y_t$ of its environment at that time, is determined by
the wave function $\psi_t$ of the system in the familiar way.

As a mathematical formula, this is completely straightforward: By
equivariance, the joint distribution of $X_t$ and $Y_t$, i.e., the
distribution of $Q_t=(X_t,Y_t)$, is $|\Psi_t(x,y)|^2$. To obtain the
conditional probability, $y$ must be replaced by $Y_t$ and the result
normalized, yielding $|\psi_t(x)|^2$ with normalized conditional wave
function $\psi_t$. 

It is also tempting to read the formula as making genuine
probability statements about real-world events, statements that are
relevant to expectations about what should actually happen. To do so, as I
shall do here, of course goes beyond simple mathematics. At the end of the
day, however, such a usage can be entirely justified  \cite{durr92}. 

I wish to focus a bit more carefully on what is suggested by the
fundamental conditional probability formula \eqref{fcpt}. I shall do so in
the next subsection, but before doing so let me rewrite the formula,
suppressing the reference to the time $t$ under consideration to obtain 
\be\label{fcp}
P(X\in dx|\,Y)=|\,\psi(x)|^2dx.
\ee
It is perhaps worthwhile to compare this with one of the fundamental
formulas of statistical mechanics, the Dobrushin-Lanford-Ruelle (DLR)
equation
\be\label{drl}
P(X\in dx|\,Y) \propto e^{-H(x|Y)/kT}dx
\ee
for the conditional distribution of the configuration of a classical system
given the configuration of its environment, a heat bath at temperature $T$.
Here $H(x|y)$, the energy of the system when its configuration is $x$,
includes the contribution to this energy arising from interaction with the
environment.  The existence of such a simple formula, which is in fact
sometimes used to define the notion of classical equilibrium state, is the
main reason that in statistical mechanics, equilibrium is so much easier to
deal with than nonequilibrium.

\subsection{Quantum Equilibrium and Absolute Uncertainty}

There are many ways that we may come to have information about a system.
It would be difficult if not impossible to consider all of the
possibilities. However, whatever the means by which the
information has been obtained, it must be reflected in a correlation
between the state of the system and suitable features of the system's
environment, such as pointer orientations, ink marks on paper, computer
printouts, or the configuration of the brain of the experimenter. All such
features are determined by the much more detailed description provided by
the complete configuration $Y$ of the environment of the system, which
contains much more information than we could hope to have access to.

Nonetheless, the fundamental conditional probability formula \eqref{fcp}
says that even this most detailed information can convey no more about the
system than knowledge of its wave function $\psi$, so that in a Bohmian
universe the most we could come to know about the configuration of a system
is that it has the quantum equilibrium distribution $\psisq$. Thus in a
Bohmian universe we have an {\em absolute uncertainty}, in the sense that
the limitations on our possible knowledge of the state of a system
expressed by \eqref{fcp} can't be overcome by any clever innovation,
regardless of whether it employs current technology or technological
breakthroughs of the distant future.

In other words, the fundamental conditional probability formula \eqref{fcp}
is a sharp expression of the inaccessibility in a Bohmian universe of
micro-reality,  of the unattainability of knowledge of the
configuration of a system that transcends the limits set by its wave
function $\psi$. This makes it very natural to regard or speak of quantum
mechanics, or the wave function, as about information, since the wave
function does indeed provide optimal information about a system. At the same
time, it seems to me that our best understanding of this informational
aspect of the wave function emerges from a theory that is primarily about
the very micro-configuration that it shows to be inaccessible!

\subsection{Random Systems}
While the fundamental conditional probability formula \eqref{fcp} seems
very strong, the following stronger version, that applies to random
systems, is also true and is often useful, particularly for a careful
analysis of the empirical implications of Bohmian mechanics for the results
of a sequence of experiments performed at different times \cite{durr92}:
\be\label{fcpr}
P(X_{\sigma}\in dx\,|\,Y_{\sigma},\sigma)=|\psi_{\sigma}(x)|^2dx.
\ee

In this formula, $\sigma$ denotes a random system, i.e., a random subsystem
with configuration $X_{\sigma}$  
at a random time $T$,
\be\label{sigma}
\sigma=(\pi,T).
\ee
Here $\pi$ is a projection, defining a random splitting  
\be
q=(\pi q, \pi^{\perp}q)=(x,y).
\ee 
For a given initial universal wave function $\Psi_0$, $\sigma$ is
determined (like everything else in a Bohmian universe) by the initial
universal configuration $Q$,
\be
\sigma=\sigma(Q)= (\pi(Q),T(Q)).
\ee
Thus
\be
X_{\sigma}=\pi Q_T,\quad Y_{\sigma}=\pi^{\perp} Q_T.
\ee
More explicitly,
\be
X_{\sigma}(Q)=\pi(Q) Q_{T(Q)},\quad Y_{\sigma}(Q)=\pi(Q)^{\perp} Q_{T(Q)}.
\ee
$\psi_{\sigma}$ is defined analogously.

The formula \eqref{fcpr} holds provided the random system obeys the
measurability condition  
\be
\{\sigma=\sigma_0\}\in {\cal F}(Y_{\sigma_0}),
\ee
which expresses the requirement that the identity of the random system be
determined by its environment. See \cite{durr92} for details. With this
condition, the notion of a random system becomes roughly analogous to that
of a {\em stopping time} in the theory of Markov processes. And the random
system fundamental conditional probability formula \eqref{fcpr} then
becomes analogous to the {\em strong Markov property,} which plays a
crucial role in the rigorous analysis of these processes.

\section{The Classical Limit} 
The classical limit of Bohmian mechanics is reasonably clear
\cite{allori02}; I don't intend to enter into any details here. Rather I wish
merely to note that it would be nice to have some rigorous mathematical
results in this direction and to make two comments: 
\begin{itemize}
\item Decoherence plays a controversial role in the classical limit of
  orthodox quantum theory. It also important for a full appreciation of
  this limit for Bohmian mechanics, where in fact it is entirely
  uncontroversial and straightforward. And insofar as decoherence is
  strongly associated with measurement and observation, Bohmian mechanics
  provides a natural explanation of the apparent importance of information
  for the emergence of classical behavior.
\item Considerations related to decoherence suggest the following: {\em In
    Bohmian mechanics an observed motion, if it seems deterministic, will
    appear to be classical.} This conjecture provides an ahistorical
  explanation of why in a Bohmian world classical mechanics would be
  discovered before Bohmian mechanics: the {\em observed} deterministic
  regularities would be classical. (Of course the real explanation, not
  unrelated, is that we live on the macroscopic level, where objects behave
  classically.)
\end{itemize}

\section{The Wave Function as Nomological}
Perhaps the most significant informational aspect of the wave function is
that it is best regarded as fundamentally nomological, as a component of
physical law rather than of the physical reality described by the law
\cite{durr97,goldstein02}, as I shall now argue.

The wave function in Bohmian mechanics is rather odd in at least two
ways---how it behaves and the kind of thing that it is:
\begin{itemize}
\item While the wave function is
crucially implicated in the motion of the particles, via equation
\eqref{g}, the particles can have no effect whatsoever on the wave
function, since Schr\"odinger's equation is an autonomous equation for
$\psi$, that does not involve the configuration $Q$. 
\item For an $N$-particle system the wave function $\psi(q)= \psi(\mathbf
  q_{_1}, \dots,  \mathbf q_{_N})$ is, unlike the electromagnetic field, not
  a field on physical space but on configuration space, an 
  abstract space of great dimension. 
\end{itemize}
Though it is possible to perhaps temper these oddities with suitable
responses---for example that the action-reaction principle is normally
associated with conservation of momentum, which in turn is now taken to be
an expression of  translation invariance,  a feature of Bohmian
mechanics---I think we should take them more seriously, and try to
come to grips with what they might be telling us.  

We are familiar with an object that is somewhat similar to the wave
function, namely the Hamiltonian of classical mechanics, a function on a
space, phase space, of even higher dimension than configuration space. In
fact the classical Hamiltonian is surprisingly analogous to the wave
function, or, more precisely, to its logarithm:
\be
\text{log}\, \psi(q)\leftrightarrow H(q,p)=H(\mathscr{X})
\ee
where $\mathscr{X}=(q,p)=(\mathbf q_{_1}, \dots, \mathbf q_{_N},\mathbf p_{_1},
\dots, \mathbf p_{_N})$ is the phase space variable.
Corresponding to these objects we have the respective equations of motion  
\be
dQ/dt =der(\text{log}\, \psi)\leftrightarrow d\mathscr{X}/dt=der H
\ee
with $der$ representing suitable first derivatives.  

Note as well that both $\text{log}\, \psi(q)$ and $H(\mathscr{X})$ are
defined only up to an additive constant.  For ``normalized'' choices we
further have that
\be
\text{log Prob} \propto\ \text{log} |\psi| \leftrightarrow\text{log Prob} \propto -H
\ee
(This should not be taken too seriously!)

Of course nobody has a problem with the fact that the Hamiltonian is a
function on the phase space, since it is not a dynamical variable at all
but rather an object that generates the classical Hamiltonian dynamics. 
As such, it would not be expected to be affected by anything physical
either. 

But there are some important differences between $\psi$ and $H$.
Unlike $H$, $\psi$ typically changes with time and serves moreover as (the
paradigmatic) initial condition in quantum mechanics:
\begin{itemize}
\item $\psi_t$ is dynamical.
\item $\psi$ is controllable.
\end{itemize}
These quite naturally tend to undercut the suggestion that $\psi$ should be
regarded as nomological, since, unlike dynamical variables, laws  are not
supposed to be like that. However, it is important in this regard to
bear in mind the distinction:
\be
\psi\quad \text{versus} \quad \Psi.
\ee

\subsection{The Universal Level}
In Bohmian mechanics the wave function $\Psi$ of the universe is
fundamental, while the wave function $\psi$ of a subsystem of the universe
is derivative, defined in terms of $\Psi$ by \eqref{cwf}. Thus the crucial
question about the nature of the wave function in Bohmian mechanics must
concern   $\Psi$; once this is settled  the nature of $\psi$ will then be
determined. 

Accordingly, the claim that the wave function in Bohmian mechanics is
nomological should be understood as referring primarily to the wave
function  $\Psi$ of the universe, concerning which it is important to note
the following:
\begin{itemize}
\item $\Psi$ is not controllable: it is what it is.
\item If we are seriously considering the universal or cosmological level,
  then we should perhaps take the lessons of general relativity into
  account. Now the significance of being ``dynamical,'' of having an
  explicit time dependence, is transformed by general relativity, and indeed
  by special relativity, since the (3,1) splitting of space and time is
  thereby transformed to a 3 + 1 = 4 dimensional space-time that admits no
  special splitting.
\item There may well be no ``$t$'' in $\Psi$. The Wheeler-DeWitt equation,
  the most famous equation for the wave function of the universe in quantum
  gravity, is 
  of the form
\be \label{wd}
\mathscr{H}\Psi=0
\ee
with $\mathscr{H}$ a sort of Laplacian on a space of configurations of
suitable structures on a 3-dimensional space and with $\Psi$ a function on
that configuration space that does not contain a time variable at all. For
orthodox quantum theory this is a problem, the {\em problem of time}: of
how change can arise when the wave function does not change. But for
Bohmian mechanics, that the wave function does not change is, far from being
a problem, just what the doctor ordered for a law, one that governs the
changes that really matter in a Bohmian universe: of the variables $Q$
describing the fundamental objects in the theory, including the 3-geometry
and matter. The evolution equation should be regarded as more or less of a
form
\be\label{G}
dQ/dt = v^{\Psi}(Q)
\ee
roughly analogous to \eqref{g}, one that
defines an evolution that is natural for the PO of the theory under
consideration. 
\end{itemize}

\subsection{Schr\"odinger's Equation as Phenomenological}
Of course, accustomed as we are to Schr\"odinger's equation, we can hardly
resist regarding the wave function as time dependent. And it is hard to
imagine a simple description of the measurement process in quantum
mechanics that does not invoke a time dependent wave function. In this
regard, it is important to bear in mind that the fact---if it is a
fact---that the wave function $\Psi$ of the universe does not change in no
way precludes the wave function $\psi$ of a subsystem from changing. On the
contrary, since a solution to the Wheeler-deWitt equation \eqref{wd} is in
fact just a special (time-independent) solution to Schr\"odinger's
equation, it follows, as said earlier in Section 4---assuming that the
considerations alluded to earlier for Bohmian mechanics apply to the
relevant generalization of Bohmian mechanics---that the conditional wave
function
\be
\psi_t(x)=\Psi(x,Y_t)
\ee
will {evolve} according to Schr\"odinger's equation when the
subsystem is suitably decoupled from its environment (and $\mathscr{H}$ is
of the appropriate form). 

In this way what is widely taken to be the fundamental equation of quantum
mechanics, the time-dependent Schr\"odinger equation, might turn out to be
merely phenomenological: an emergent equation for the wave function of
suitable subsystems of a Bohmian universe. Moreover, even the
time-independent Schr\"odinger equation \eqref{wd} might best be regarded
accidental rather than fundamental. What happens in a Bohmian universe with
universal wave function $\Psi$ is entirely determined by the equation of
motion \eqref{G} for the PO of the theory. This theory is then determined
by $\Psi$ and the form of $v^{\Psi}$. \eqref{wd} will be fundamental only
if it constrains the choice of $\Psi$, but this need not be so. It might
well be that the choice of $\Psi$ is fundamentally constrained by entirely
different considerations, such as the desired symmetry properties of the
resulting theory, with the fact that $\Psi$ also obeys \eqref{wd} thus
being accidental.

\subsection{Two Transitions}
Suppose what I've written here about the fundamental Bohmian mechanics,
Universal Bohmian Mechanics (UBM), is correct. Then our understanding of
the nature of quantum reality is completely transformed, as is the question
about the nature of the wave function in quantum mechanics with which we
began:\newpage
\begin{itemize}
\item
\quad\quad\quad
\parbox{1in}{\bc OQT\\$\Psi$\ec}$\longrightarrow$\parbox{1in}{\bc BM\\$(\Psi,Q)$\ec}$\longrightarrow$\parbox{1in}{\bc UBM\\$Q$\ec}

\item

\parbox{2in}{\bc?\\
?$\quad\psi\quad$?\\
?\ec}
$\ \longrightarrow$
\ \ \parbox{2in}{\bc?\\
?$\quad\Psi\quad$?\\
?\ec}

\end{itemize}
The first transition is of the basic variables involved as we proceed from
orthodox quantum theory, which seems to many to involve as a basic variable
only the wave function $\Psi$---and certainly no hidden variables; to the
usual Bohmian mechanics, whose basic variables are $\Psi$ and $Q$; to UBM,
with $Q$ the  only fundamental physical variable, the
universal wave function $\Psi$ remaining only as a mathematical object
convenient for expressing the law of motion \eqref{G}. 

And accordingly, the question about the meaning of the wave function in
quantum mechanics is utterly transformed, from something like, What on
earth does the wave function $\psi$ of a system physically describe? to,
Why on earth  should a wave function $\Psi$ play a prominent role in the
law of motion \eqref{G} defining quantum theory? What's so good about such
a motion?

Once we recognize that the wave function is nomological we are confronted
with a transformed landscape for understanding why nature should be quantum
mechanical. We will fully comprehend this once we understand what is so
special and compelling about a motion governed by a wave function in
Bohmian way.

\subsection{Nomological versus Nonnomological}\label{NvN}
I can well imagine many physicists, when confronted with the question of
whether the wave function should be regarded as nomological or as more
concretely physical, responding with a loud, Who cares! What difference
does it make? But quite aside from the fact that it is conceptually
valuable to understand the nature of the objects we are dealing with in a
fundamental physical theory, the question matters in a practical way. It is
relevant to our expectations for future theoretical developments.

In particular, laws should be simple, so that if $\Psi$ is nomological, it
too---and the law of motion \eqref{G} it defines--should somehow be simple
as well. The contention that $\Psi$ is nomological would be severely
undermined if this were not achievable.

Simplicity of course comes in many varieties. $\Psi$ might be
straightforwardly simple, i.e., a simple function of its argument,
expressible in a compelling way using the structure at hand. It might be
simple because it is a solution, perhaps the unique solution, to a simple
equation. Or it might be the case that there is a compelling principle, one
that is simple and elegant, that is satisfied, perhaps uniquely, by a law
of motion of the form \eqref{G} with a specific $\Psi$ and $v^{\Psi}$. For
example, the principle might express a very strong symmetry condition.

\subsection{ Covariant Geometrodynamics}\label{cg}
Stefan Teufel and I have examined such a possibility for quantum gravity
\cite{goldstein991}, with the symmetry principle that of 4-diffeomorphism
invariance.  Within (an extension of) the framework of the ADM formalism,
the dynamical formulation of general relativity of Arnowitt, Deser, and
Misner
\cite{arnowitt}, we considered the possibilities for a first-order covariant
geometrodynamics.

In the ADM formalism the dynamics corresponds to the change of structures,
most importantly a 3-geometry, on a space-like hypersurface as that surface
is infinitesimally deformed. In a theory for which there is no special
foliation of space-time into hypersurfaces (that might define the notion of
simultaneity if it existed), a hypersurface $\Sigma$ can naturally be
deformed in an infinite dimensional variety of ways. These are given by the
function $\text{N}=(N,\vec{N})$, where $N=N(x),\ x\in \Sigma$, is the {\em
  lapse} function describing deformations normal to the surface, and
$\vec{N}=\vec{N}(x)$ is the {\em shift} function describing deformations in
the surface, i.e., infinitesimal 3-diffeomorphisms.  Corresponding to the many
possible deformations N, one often speaks here of a {\em
  multi-fingered time}.

The deformations N form an algebra, the Dirac Algebra, which is almost a
Lie algebra and should be regarded as somehow corresponding to the group of
4-diffeomorphisms of space-time. The Dirac Algebra, with Dirac bracket
$[\text{N},\text{M}]$, is defined, using linearity, by 
\be
[N,M] = N\vec{\nabla}M - M\vec{\nabla}N;\quad
[N,\vec{M}]=\vec{M}\cdot\vec{\nabla}N
\ee
together with the usual Lie bracket $[\vec{N},\vec{M}]$ for the Lie algebra
of the group of 3-diffeomorphisms. 

Within this multi-fingered time framework, a first-order dynamics
corresponds, not to a single vector field on the configuration space
$\mathscr{Q}$---of decorations of $\Sigma$---in which the evolution occurs,
but to a choice of vector field $\mathscr{V}(\text{N})$ for each
deformation N. (See \cite{hojman} for the more familiar second-order, phase
space, Poisson bracket approach.) Moreover, it seems, at least
heuristically, that the dynamics so defined will be covariant precisely in
case $\mathscr{V}(\text{N})$ forms a {\em representation of the Dirac
  algebra}:
\be\label{DA}
[\mathscr{V}(\text{N}),\mathscr{V}(\text{M})]=\mathscr{V}([\text{N},\text{M}]),
\ee
where the bracket on the left is the Lie bracket of vector fields. 

The claim that such a dynamics is covariant is intended to convey that it
defines a 4-diffeomorphism invariant law for a decoration of space-time; a
crucial ingredient in this is that the dynamics be path-independent: that
two different foliations that connect the same pair $\Sigma_i$ and
$\Sigma_f$ of hypersurfaces, corresponding to two different paths through
the multi-fingered time \{N\}, yield the same evolution map connecting
decorations of $\Sigma_i$ to decorations of $\Sigma_f$.

The requirement that $\mathscr{V}(\text{N})$ form a representation of the
Dirac Algebra is a very strong symmetry condition. Our hope was that it was
so strong that it would force the dynamics to be quantum mechanical:
$\mathscr{V}(\text{N})=\mathscr{V}^{\Psi}(\text{N})$ where
$\mathscr{V}^{\Psi}$ is a suitable functional of $\Psi$, with $\Psi$ obeying an
equation of the form \eqref{wd}. It seems, however, for pure quantum
gravity, with $\mathscr{Q}$ the space of 3-geometries (super-space), that any
covariant dynamics is classical, yielding 4-geometries that obey the
Einstein equations, with a possible cosmological constant, and
with no genuinely quantum mechanical possibilities arising.

When, in addition to geometry, structures corresponding to matter are
included in $\mathscr{Q}$, it is not at all clear what the possibilities
are for the representations of the Dirac algebra. It seems a long shot that
a quantum mechanical dynamics could be selected in this way as the only
possibility, let alone one that corresponds to a more or less unique
$\Psi$. But since a positive result in this direction would be so exciting,
this program seems well worth pursuing further---even if only to establish
its impossibility.

\subsection{The Value of Principle}
It is often suggested that what is unsatisfactory about orthodox quantum
theory is that it was not formulated as a theory based on a compelling
principle, an information theoretic principle or whatever. Often such a
derivation is then supplied. If, as is usually the case, what we then
arrive at is---as presumably intended---plain old orthodox quantum theory, I
find myself unsatisfied by the accomplishment. 

The reason is this. The problem with orthodox quantum theory is not that
the principles from which it might be derived are unclear or absent, but
that the theory itself is, in the words of Bell \cite[page 173]{bell87},
``unprofessionally vague and ambiguous.'' Thus if derivation from a
principle only yields orthodox quantum theory, how has the problem of
understanding what quantum mechanics actually says been at all addressed?
Of course, if the derivation yields, not orthodox quantum theory, but an
improved formulation of quantum mechanics, then the problem may well have
been alleviated. But this rarely happens.

It is fine and good to want to understand why a theory should hold. But
before worrying about this we should first get clear about what the theory
in fact says. The crucial distinction is between the question, Why?  and
the question, What?: Why should quantum theory hold?  versus What does
quantum theory say? A derivation of quantum theory will address the real
problem with quantum mechanics if it provides answer to What? and
not just an answer to Why? The sorts of derivation from a principle
contemplated in Sections \ref{NvN} and \ref{cg} are of this form.

\section{Quantum Rationality}
I conclude with two quotations. The first addresses the question, if
Bohmian mechanics is so simple and elegant, and accounts for quantum
phenomena is such a straightforward way, why is this not recognized by the
physics community? 
\bq
I  know that most men, including those at ease with problems of the highest
complexity, can seldom accept even the simplest and most obvious truth if
it be such as would oblige them to admit the falsity of conclusions which
they have delighted in explaining to colleagues, which they have proudly
taught to others, and which they have woven, thread by thread, into the
fabric of their lives. ({\em Leo Tolstoy})
\eq
I have another reason for quoting Tolstoy here: I would like to know where
he said this. If any reader knows, I would be very grateful if he contacted
me with the information.

The Tolstoy is of course a bit depressing. So I will conclude on a more
optimistic note \cite[page 145]{lakatos}, from the philosopher of science Imre Lakatos, who was an
early teacher of Jeffrey's.
\bq
In the new, post-1925 quantum theory the `anarchist' position became
dominant and modern quantum physics, in its `Copenhagen interpretation',
became one of the main standard bearers of  philosophical obscurantism. In
the {\em new} theory Bohr's notorious `complementarity principle' enthroned
[weak] inconsistency as a basic ultimate feature of nature, and merged
subjectivist positivism and antilogical dialectic and even ordinary
language philosophy into one unholy alliance. After 1925 Bohr and his
associates introduced a new and unprecedented lowering of critical
standards for scientific theories. This led to a defeat of reason within
modern physics and to an anarchist cult of incomprehensible
chaos. ({\em 1965})
\eq

\section{Acknowledgements}
I am grateful to Michael Kiessling, Roderich Tumulka, and Nino Zangh\`\i\
for their help. This work was supported in part by NSF Grant DMS--0504504.


\begin{thebibliography}{10}

\bibitem{einstein49}
{A.\ Einstein, ``Reply to Criticisms," in {\em Albert Einstein,
  Philosopher-Scientist}, ed. P.A.\ Schilpp, (The Library of Living
  Philosophers, 1949).}

\bibitem{bell87}
{J.S.\ Bell, {\em Speakable and unspeakable in quantum mechanics}, (Cambridge
  University Press, Cambridge, 1987).}

\bibitem{debroglie28}
{L.\ de Broglie, in {\em \'Electrons et Photons: Rapports et Discussions du
  Cinqui\`eme Conseil de Physique}, ed.\ J.\ Bordet, (Gauthier-Villars, Paris,
  1928) 105; English translation: G.\ Bacciagaluppi and A.\ Valentini, {\em
  Quantum Theory at the Crossroads}, (Cambridge University Press,
  forthcoming).}

\bibitem{bohm521}
{D.\ Bohm, {\em Phys.\ Rev.}\ {\bf 85}, 166 (1952).}

\bibitem{bohm93}
{D.\ Bohm and B.J.\ Hiley, {\em The Undivided Universe}, (Routledge, New York,
  1993).}

\bibitem{weihs}
{G.\ Weihs, {\em Nature}\ {\bf 445}, 723 (2007).}

\bibitem{griffiths}
{D.J.\ Griffiths, {\em Introduction to Quantum Mechanics (2nd Edition)},
  (Benjamin Cummings, 2004).}

\bibitem{britannica}
{{\bf quantum mechanics}, in {\em Encyclopædia Britannica} (2007), retrieved
  from Encyclopædia Britannica Online, 12 June 2007,
  http://www.britannica.com/eb/article-77521.}

\bibitem{allori07}
{V.\ Allori, S.\ Goldstein, R.\ Tumulka and N.\ Zangh\`\i, {\em Brit. J. Phil.
  Sci.}, to appear (2007) and quant-ph/0603027.}

\bibitem{philippidis79}
{C.\ Philippidis, C.\ Dewdney and B.J.\ Hiley, {\em Il Nuovo Cimento} {\bf 52},
  15 (1979)}.

\bibitem{bohm522}
{D.\ Bohm, {\em Phys.\ Rev.}\ {\bf 85}, 180 (1952).}

\bibitem{durr92}
{D.\ D\"urr, S.\ Goldstein and N.\ Zangh\`\i, {\em J.\ Stat.\ Phys.}\ {\bf 67},
  843 (1992) and quant-ph/0308039.}

\bibitem{durr061}
{D. D\"urr, S. Goldstein, T. Moser and N. Zangh\`\i, {\em Commun. Math. Phys.}
  {\bf 266}, 665 (2006).}

\bibitem{durr041}
{D.\ D\"urr, S.\ Goldstein and N.\ Zangh\`\i, {\em J.\ Stat.\ Phys.}\ {\bf
  116}, 959 (2004) and quant-ph/0308038.}

\bibitem{allori02}
{V.\ Allori, D.\ D\"urr, S.\ Goldstein and N.\ Zangh\`\i, {\em Journal of
  Optics B} {\bf 4}, 482 (2002) and quant-ph/0112005}.

\bibitem{goldstein07}
{S.\ Goldstein and W.\ Struyve, {\em J.\ Stat.\ Phys.}, to appear (2007) and
  0704.3070 [quant-ph].}

\bibitem{durr97}
{D.\ D\"urr, S.\ Goldstein and N.\ Zangh\`\i, ``Bohmian Mechanics and the
  Meaning of the Wave Function," in {\em Experimental Metaphysics---Quantum
  Mechanical Studies for Abner Shimony, Volume One; Boston Studies in the
  Philosophy of Science} {\bf 193}, eds. R.S.\ Cohen, M.\ Horne and J.\
  Stachel, (Kluwer Academic Publishers, Boston, 1997) and quant-ph/9512031.}

\bibitem{goldstein02}
{S.\ Goldstein and S.\ Teufel, ``Quantum Spacetime without Observers:
  Ontological Clarity and the Conceptual Foundations of Quantum Gravity," in
  {\em Physics meets Philosophy at the Planck Scale}, eds. C. Callender and N.
  Huggett, (Cambridge University Press, 2001) and quant-ph/9902018.}

\bibitem{goldstein991}
{S.\ Goldstein and S. Teufel, ``Covariant Geometrodynamics and Bohmian Quantum
  Gravity," (1999), unpublished preprint.}

\bibitem{arnowitt}
{R.\ Arnowitt, S.\ Deser and C.W.\ Misner, ``The dynamics of general
  relativity," in {\em Gravitation: An Introduction to Current Research}, ed.
  L. Witten, (Wiley, New York, 1962).}

\bibitem{hojman}
{S.A.\ Hojman, K.\ Kucha\v{r} and C.\ Teitelboim, {\em Annals of Physics} {\bf
  96\/}, 88 (1976).}

\bibitem{lakatos}
{I.\ Lakatos, ``Falsification and the Methodology of Scientific Research
  Programmes," in {\em Criticism and the Growth of Knowledge}, eds. I.\ Lakatos
  and A. Musgrave, (Cambridge University Press, 1970).}

\end{thebibliography}
\end{document}